\PassOptionsToPackage{table}{xcolor}
\documentclass[reprint,aps,prl,floatfix]{revtex4-2}
\usepackage{bm}
\usepackage{amssymb}
\usepackage{amsmath}
\usepackage{graphicx}
\usepackage{rotating}
\usepackage{epsfig}
\usepackage{psfrag}
\usepackage{amsmath}
\usepackage{subfigure}
\usepackage{braket}
\usepackage{tikz}
\usepackage{stmaryrd}
\usepackage{wasysym}
\usepackage{array}
\usepackage{booktabs}
\usepackage{colortbl}
\usepackage[table]{xcolor}

\newcommand{\bea}{\begin{eqnarray}}
\newcommand{\eea}{\end{eqnarray}}
\newcommand{\bpm}{\begin{pmatrix}}
\newcommand{\epm}{\end{pmatrix}}

\newcommand{\cmark}{\text{\checkmark}}
\newcommand{\xmark}{\text{$\mathbf{\times}$}}

\usepackage[unicode=true,colorlinks=true,pdfpagemode=UseOutlines,pdfstartview=Fith]{hyperref}
\hypersetup{linkcolor=blue,citecolor=blue,urlcolor=blue}

\begin{document}
\title{Altermagnets with topological order in Kitaev bilayers}

\author{Aayush Vijayvargia, Ezra Day-Roberts, Antia S. Botana, Onur Erten}
\affiliation{Department of Physics, Arizona State University, Tempe, AZ 85287, USA}

\begin{abstract}
Building on recent advancements in altermagnetism, we develop a highly-frustrated magnetic model with Kitaev-like interactions that integrates key aspects of both quantum spin liquids and altermagnets. While the ground state is a gapless quantum spin liquid, our analysis indicates that an altermagnetic local order emerges upon the introduction of additional interactions that gap the excitation spectrum and give rise to a $\mathbb{Z}_2 $ topological order. This magnetically-fragmented topological altermagnet has fractionalized fermionic excitations with momentum-dependent splitting, in stark contrast to both standard altermagnets and Kitaev spin liquids. In addition, we discover two more altermagnetic phases, including a pseudo-altermagnet that exhibits splitting in the absence of a local order and a half-altermagnet that possesses only one type of fractionalized excitations, similar to a half-metal. We discuss experimental approaches for detecting these phases, including layer-dependent spin and heat transport. Our results highlight the rich physics that can arise due to the interplay between altermagnetism and fractionalized excitations in quantum magnets.

\end{abstract}
\maketitle
\noindent
An unconventional form of collinear magnetic order,
named altermagnetism, has recently been introduced.  Altermagnets (AMs) exhibit momentum-dependent spin splitting in their band structures in the absence of a net moment or spin-orbit coupling, integrating traits previously considered exclusive to conventional ferromagnets and antiferromagnets \cite{vsmejkal2022beyond, vsmejkal2022emerging, mcclarty2024landau, osumi2024observation, Yang2025, Fernandes_PRB2024}. As a consequence of this unique combination of properties, AMs have great potential for low-energy spintronics applications \cite{bai2024altermagnetism}. To date, research on AMs has mostly focused on identifying families of materials with different types of spin-splitting, while treating their magnetism semi-classically \cite{wei2024crystal}. In return, the magnetic order acts as a static (classical) alternating on-site Zeeman term on the electronic quasiparticles. Integrating key aspects of frustrated quantum and altermagnetism is, however, still a largely unexplored area \cite{Sobral_arXiv2024}. In the context of frustrated magnetism, Kitaev's honeycomb model is foundational since it is the first exactly solvable model with a spin liquid ground state exhibiting both Abelian and non-Abelian anyons \cite{Kitaev_AnnPhys2006}. %Kitaev model can be generalized to other tri-coordinated ($z=3$) lattices such as hyperhoneycomb, hyperoctagon and square-octagon lattices\cite{Lee_PRB2014, Mishchenko_PRB2017, Jahin_PRA2022, Yamada_PRR2021, Li_PRR2023, Baskaran_arxiv2009,Lukas_PRB2024,Lai_PRB2011,Yang_PRB2007}.

\begin{figure}[t]
    \centering
    \includegraphics[width=\linewidth]{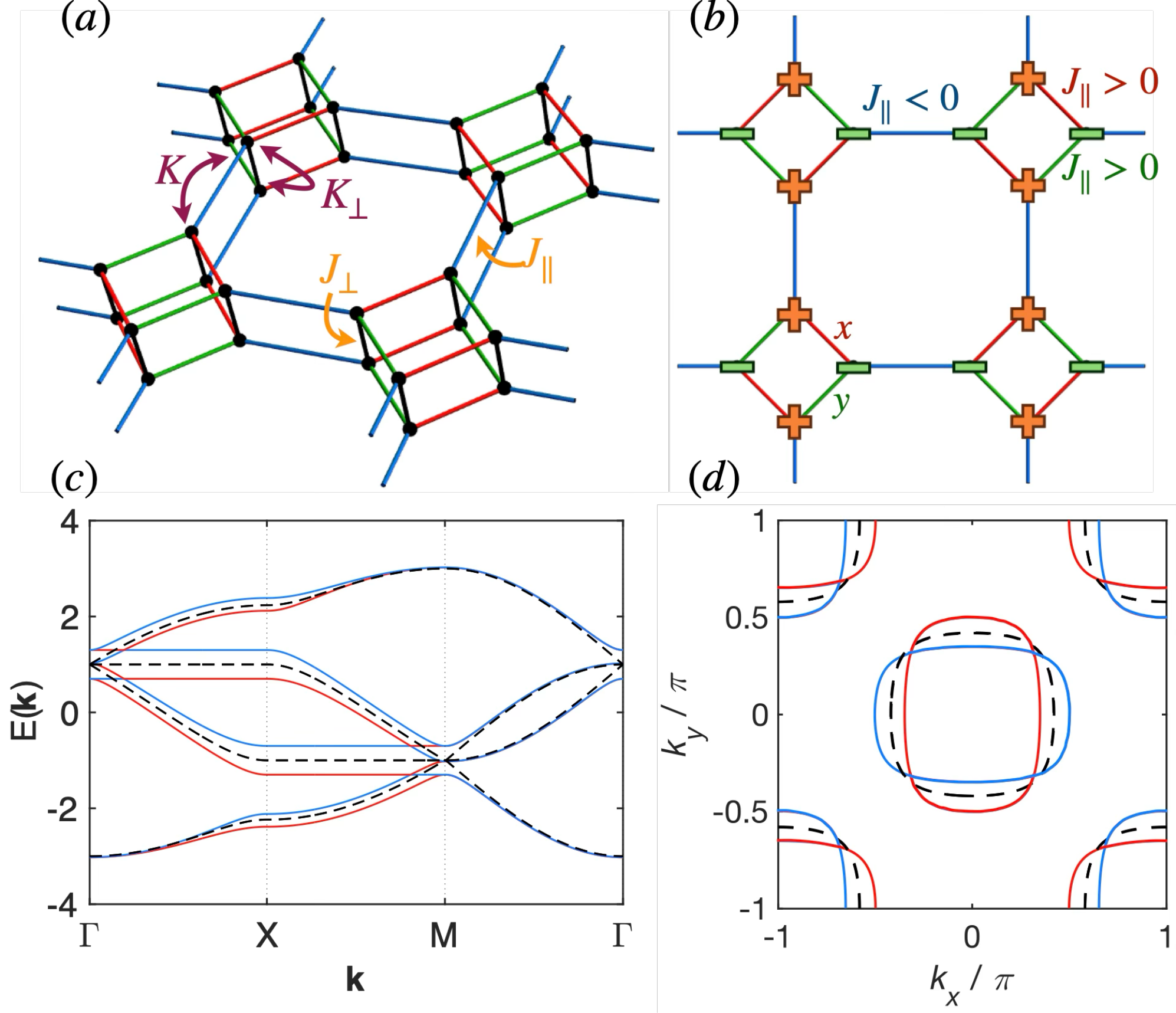}
    \caption{(a) Schematic of our model defined on bilayer square-octagon lattice. $K, K_\perp, J_\perp$ and $J_\parallel$ are four different coupling terms in our model (eq.~\ref{eq:1} and~\ref{eq:2}) (b) Top-view of the bilayer with $'+'$and $'-'$ symbols indicating the altermagnetic magnetization alongside the signs of $J_\parallel$ coupling. (c) Excitation spectrum along the high-symmetry points in the Brillouin Zone and (d) Fermi surface obtained in the zero-flux sector of the model. Red and blue lines show altermagnetic splitting while the dashed line is for no AM order.}
    \label{fig:1}
\end{figure}

In this letter, we construct a quantum spin model where we treat the emergence of AM order beyond the semi-classical approximation. Our model is based on the generalizations of the Kitaev's honeycomb model \cite{Kitaev_AnnPhys2006} defined on a bilayer square-octagon lattice as depicted in Fig.~\ref{fig:1}(a). We examine the square-octagon model not only because it exhibits interesting spin liquid phases wherein Majorana Fermi surfaces can be stabilized \cite{Baskaran_arxiv2009, Lai_PRB2011, Yamada_PRR2021, Li_PRR2023, Lukas_PRB2024, Yang_PRB2007} but also because its diamond-shaped primitive unit cell allows $d$-wave altermagnetic order \cite{Bose_PRB2024}. Unlike a standard AM, the ground state exhibits $\mathbb{Z}_2$ topological order, and two bands of fractionalized excitations that are split in momentum space (Fig.~\ref{fig:1}(c)-(d)). In addition, we discover two new phases: a pseudo-altermagnet that exhibits a momentum-dependent splitting in the excitation spectrum without a local order and a half-altermagnet that has only a single band of excitations which are distorted due to the altermagnetic ordering.

{\it General considerations and microscopic model.}
We start by introducing some broad principles of our magnetic model and the requirements to realize AM in it. In the Kitaev model, spin-$1/2$ degrees of freedom (DOF) are fractionalized by introducing four Majorana fermions per site, $\sigma^\alpha = i b^\alpha c$ ($\alpha = x, y, z)$. While $c$ Majorana fermions delocalize and form spinon bands, the $b^\alpha$ Majorana fermions are localized on bonds and form static $\mathbb{Z}_2$ fluxes. A single band of spinons in the Kitaev model cannot exhibit AM splitting as it is non-degenerate. Even a bilayer generalization of the Kitaev model with two Majorana bands is not sufficient since the onsite Zeeman terms hybridize the Majorana fermions, instead of shifting their energy levels \cite{Seifert_PRb2018, Vijayvargia_PRB2024}. Moreover, a bilayer setup breaks the integrability of the Kitaev model since the interlayer exchange terms do not commute with intralayer fluxes, preventing its exact solution in addition to destabilizing the spin liquid \cite{Tomishige_PRB2018, Tomishige_PRB2019}. Therefore, we deduce that a minimum of two complex fermions (or four Majorana fermions) per site is required to observe AM splitting. 

One way to increase the DOF is to extend the model from spin-$1/2$ to spin-$3/2$. The spin-$3/2$ operators can be expressed in terms of four-dimensional $\Gamma$ matrices that obey the Clifford algebra $\{\Gamma^\alpha,\Gamma^\beta\}=2\delta_{\alpha \beta}$. There are five anticommuting $\Gamma^\alpha$ matrices; ten $\Gamma^{\alpha \beta} =\frac{i}{2}[\Gamma^\alpha, \Gamma^\beta]$ and an identity operator that span the Hilbert space. These models, commonly referred to as the $\Gamma$-matrix generalizations of the Kitaev model \cite{Wu_PRB2009, Yao_PRL2009, Carvalho_PRB2018, Chulliparambil_PRB2020, Seifert_PRL2020, Natori_PRL2020, Chulliparambil_PRB2021, Nica_npjQM2023, Vijayvargia_PRR2023, Akram_PRB2023, Keskiner_PRB2023, Majumder_PRB2024, Poliakov_PRB2024}, allow for additional DOF since the $\Gamma$ matrices can be represented by six Majorana fermions, $\Gamma^\alpha = ib^\alpha c$, ($\alpha = 1,2,3,4,5$) and $\Gamma^{\alpha \beta}_i=ib_i^{\alpha} b^{\beta}_i$ (compared to four Majorana fermions for spin-$1/2$). In a bilayer square-octagon lattice, four of the six Majorana fermions constitute bond operators (including the interlayer bond, $z=4$), leaving two free Majorana fermions per site. Combining the DOF on two layers, we satisfy the requirement of two complex fermions necessary to achieve AM splitting. Thus, we consider the following Hamiltonian, $H= H_K+H_J$ where $H_K$ is an exactly solvable model with Kitaev-like interactions,

\begin{align}
    H_K= &-K\sum_{\langle ij \rangle,\nu}(\Gamma_{\nu i}^{\alpha}\Gamma_{\nu j}^{\alpha}+\Gamma_{\nu i}^{\alpha 5}\Gamma_{\nu j}^{\alpha 5})\nonumber \\
    &-K^\perp\sum_i(\Gamma_{Ai}^4\Gamma_{Bi}^4+\Gamma_{Ai}^{45}\Gamma_{Bi}^{45}) 
    \label{eq:1}
\end{align}
$\nu=A, B$ is the layer index, $\alpha = 1,2,3$ represents the three types of intralayer bonds depicted in different colors in Fig.~\ref{fig:1}(a). Similar to the Kitaev model, there are plaquette operators that commute with $H_K$. In particular, there exist two types of intralayer operators, $W_{p,S}$ and $W_{p,O}$ for square and octagon plaquettes, and three types of interlayer plaquettes, $I^\gamma_i$, where $\gamma=1,2,3$ corresponds to red, green, and blue bonds emanating from site $i$. These plaquette operators can be obtained by creating Wilson loops by multiplying $\Gamma_i^\alpha \Gamma_j^ \alpha$ around closed loops. For instance, $W_{p,S}=\prod_p\Gamma_i^\alpha \Gamma_j^ \alpha $, where the product runs over all bonds $\langle ij\rangle \in p$. The ground state of $H_K$ alone is a gapless $\mathbb{Z}_2$ spin liquid, as discussed below. To induce altermagnetic order, we include an additional term in the Hamiltonian, 
\begin{figure}[t]
    \centering
    \includegraphics[width=\linewidth]{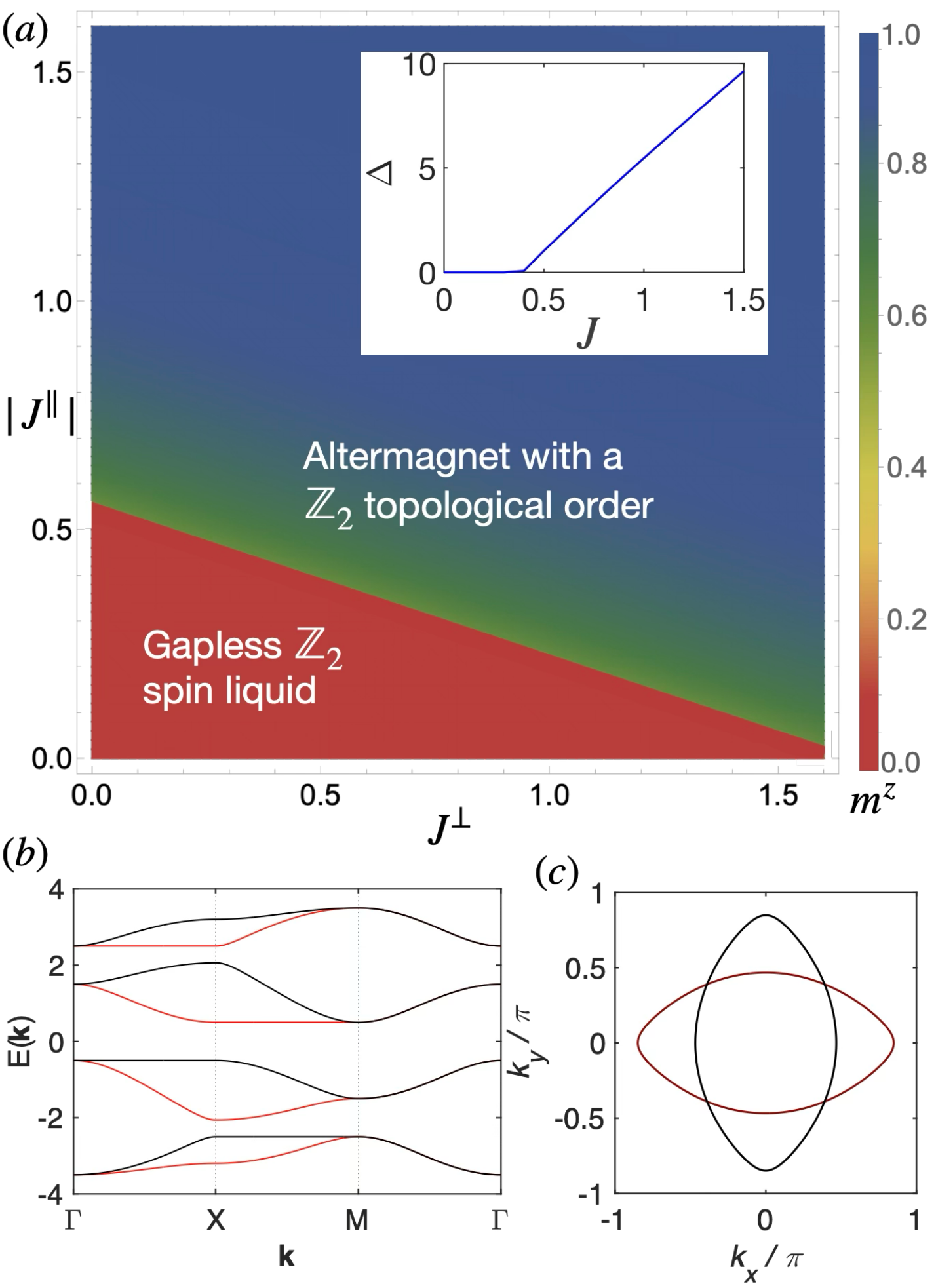}
    \caption{(a) Phase diagram of the full model with the local magnetization depicted by the color contour plot, (inset) band gap plotted as a function of $J^\perp=|J^\parallel|=J$. (b) Band structure depicting the momentum dependent splitting of the fractionalized excitations for $J^\perp=J^\parallel=0.5$. (c) Constant energy energy contour depicting the split bands at $E=-3.2$. }
    \label{fig:2}
\end{figure}
\begin{align}
    H_J = J^\perp\sum_i\Gamma_{Ai}^5\Gamma_{Bi}^5+\sum_{\langle ij \rangle,\nu}J^{\parallel}_{ ij}\Gamma_{\nu i}^5\Gamma_{\nu j}^5
    \label{eq:2}
\end{align}
Note that $H_J$ commutes with all of the intra and interlayer flux operators, maintaining the gauge structure of the full model. The Hamiltonian can be alternatively expressed in terms of spin $(\sigma)$ and orbital $(\tau)$ Pauli matrices using the relation $\Gamma^{\alpha}=-\sigma^y\otimes \tau^{\alpha}$, $\Gamma^4=\sigma^x\otimes\mathbb{I}_2$ and $\Gamma^5=-\sigma^z\otimes\mathbb{I}_2$,
\begin{align}
    H=&-\sum_{\langle ij \rangle,\nu }K(\sigma^x_{\nu i}\sigma^x_{\nu j} +\sigma^y_{\nu i}\sigma^y_{\nu j})(\tau_{\nu i}^{\alpha}\tau_{\nu j}^{\alpha})+J^{\parallel}_{ ij} \sigma^{z}_{\nu i}\sigma^{z}_{\nu j}\nonumber \\&-\sum_iK^\perp(\sigma^{x}_{Ai}\sigma^{x}_{Bi}+\sigma^{y}_{Ai}\sigma^{y}_{Bi})+J^\perp\sigma^{z}_{Ai}\sigma^{z}_{Bi}
    & \label{eq:spin}
\end{align}
Next, we introduce the Majorana fermions and relabel $b_i^4\rightarrow c_i^z$, $b_i^5\rightarrow c_i^x$ and $c_i\rightarrow c_i^y$ for notational convenience and obtain,
\begin{align}
    H=&\sum_{\langle ij\rangle,\nu}\text{i} K u_{\nu,ij}(c^x_{\nu i}c^x_{\nu j}+c^y_{\nu i}c^y_{\nu j})-J^{\parallel}_{ ij} c^x_{\nu i}c^y_{1\nu i}c^x_{\nu j}c^y_{\nu j} \nonumber \\ &+\sum_i \text{i}K^\perp u^z_i(c^x_{Ai}c^x_{Bi}+c^y_{Ai}c^y_{Bi})-J^\perp c^x_{Ai}c^y_{Ai}c^x_{Bi}c^y_{Bi}     
\end{align}
where $u_{\nu,ij}=ib_{\nu i}b_{\nu j}$ and $u^z_i=ic^z_{Ai}c^z_{Bi}$ denote the intra and interlayer bond operators, all of which commute with the Hamiltonian. Additionally, the plaquette operators can be expressed as products of bond operators.
%Note that the Majorana representation illustrates the need to have an inter-layer $K_\perp$ term in the non-interacting Hamiltonian: The $c^z$ Majorana fermion in the absence of the $K_\perp$ term remains flat at the Fermi level as there is no other term in the Hamiltonian that couples with $c^z$. Therefore, we introduce the $K_\perp$ term to gap out the Majorana band. We fix $K_\perp=10^{-3}$, as a small gap is sufficient for our calculations.
%Plaquette operators that are the product of the bond operators along a plaquette are constants of motion. There are two types of intra-layer plaquette operators in our model: $S_p = \prod_{\langle ij \rangle \in p} u_{\nu,ij}, \ O_p = \prod_{\langle ij \rangle \in p} u_{\nu,ij} $ where $S_p$ and $O_p$ denote squares and octagon plaquettes. As the number of squares and octagons in a square-octagon lattice is equal, they can be labeled by a single label: $p$. In addition, the inter-layer plaquette operators, $L_{ij}=u_{1,ij}u^z_iu_{2,ij}u^z_j$ are the product of the bond variables of a plaquette involving bonds of the two layers and the bond variables connecting the two layers.  These plaquette operators commute with both $H_K$, and $H_J$ consequently, the flux is a good quantum number.
The Majorana representation is redundant and the physical states must be restricted to the eigenstates of $D_{\nu j}=ib^1_{\nu j}b^2_{\nu j}b^3_{\nu j}c^z_{\nu j}c^x_{\nu j}c^y_{\nu j}$, with eigenvalue 1. These constraints are enforced by the projection operator $P_\nu=\prod_i(1+D_{\nu i})/2$. It is important to note that while the plaquette eigenvalues are still good quantum numbers (constants of motion), the model is no longer integrible, as the $H_J$ terms do not simplify to bilinear forms in Majorana fermions. 

%The bond operators, $u_{\nu, ij}$ and $u^z_i$ commute with $H$ and therefore are conserved, with eigenvalues $\pm 1$. A $\mathbb{Z}_{2}$ gauge transformation at site $i$ for layer $\nu$ involves flipping the signs of the Majorana fermions and bond operators, $c_{\nu i}^{\alpha} \rightarrow -c_{\nu i}^{\alpha};~ u_{\nu, \braket{ij}} \rightarrow -u_{\nu, \braket{ij}}$. This implies for a given flux sector, a gauge can be chosen by fixing all the bond operator eigenvalues: $u_{\nu, \braket{ij}}$ and $u^z_i$.

Next, we combine two Majorana fermions per site to form a Dirac fermion:
$f_{A(B) i}=(c^x_{A(B) i}-ic^y_{A(B) i})/2$. 
\begin{eqnarray}
    &&H=\sum_{\langle ij \rangle,\nu}2K u_{\nu,ij}(f_{\nu i}^{\dag}f_{\nu j}+ {\rm H.c.})+J^{\parallel}_{ij} (n_{\nu i}+n_{\nu j} -1)^2  \nonumber 
    \\ &&+\sum_{i}2K^\perp u^z_i(f_{A i}^{\dag}f_{B i}+ {\rm H.c.})+J^\perp(n_{A i}+n_{B i} -1)^2   
    \label{eq:fHam}
\end{eqnarray}
%\begin{align}
 %   H=& \sum_{\langle ij \rangle,\nu}2K u_{\nu,ij}(f_{\nu i}^{\dag}f_{\nu j}+ {\rm H.c.})+J^{\parallel}_{ij} (n_{\nu i}+n_{\nu j} -1)^2  \nonumber 
  %  \\ &+\sum_{i}2K^\perp u^z_i(f_{A i}^{\dag}f_{B i}+ {\rm H.c.})+J^\perp(n_{A i}+n_{B i} -1)^2   
   % \end{align}
where $n_{\nu i}=f^\dag_{\nu i}f_{\nu i}$. We implemented a gauge transformation to make all the hopping coefficients real (see Supplementary Material (SM) for details). Also note that $n_{i}$ is related to $S_{i}^z$ polarization since $\sigma^z_{i} = (2n_{i}-1)$.

{\it The exact solution of $H_K$.} We first discuss the properties of the ground state of $H_K$ in the absence of $H_J$. According to Lieb's theorem \cite{Lieb1994}, the ground state of the monolayer Hamiltonian lies in the $\pi$-flux sector for both square and octagon plaquettes. The ground state in this case is gapped and remains so for $K^\perp/K \ll 1$. However, since our objective is to induce an altermagnetic ordering via $H_J$, it is crucial to have a finite density of states at the chemical potential. Ref.~\citenum{Baskaran_arxiv2009} pointed out that the zero-flux sector of the monolayer model is gapless, hosting a Majorana Fermi surface as shown in Fig.~\ref{fig:1}(d) (dashed lines).  To shift the ground state to the zero-flux sector, we introduce a term that couples to the plaquette operators, $-\lambda(\sum_p(W_{p,S}+W_{p,O})+ \sum_{i,\alpha} I_i^\alpha)$. This term does not alter the eigenstates of the Hamiltonian but shifts the flux gap. For $\lambda/K\gg1$, it ensures that the zero-flux sector becomes the minimum energy sector. In this flux sector, all $u_{\nu,ij}$ and $u^z_i$ can be fixed to $1$\cite{Baskaran_arxiv2009}. 

{\it Emergence of a topological altermagnet.} Next, we examine the effects of $H_J$ on the ground state of $H_K$. We choose an antiferromagnetic interlayer Ising interaction, ($J^\perp>0$), and intralayer Ising interaction: antiferromagnetic $J_{ij}^\parallel>0$ for red and green bonds within a square and a ferromagnetic $J_{ij}^\parallel<0$ for blue bonds connecting nearest-neighbor squares as shown in Fig. \ref{fig:1}(b). This choice naturally stabilizes a bilayer version of AM ordering described in Ref.~\citenum{Bose_PRB2024}. 

Next, we decouple $H_J$ within mean-field theory and define a local order parameter $m_i^z=\langle n_{A i} - n_{B i}\rangle$. Note that $m_i^z$ is invariant under local $\mathbb{Z}_2$ gauge transformations and in terms of the original degrees of freedom, $m_i^z = \langle (\sigma_A^z-\sigma_B^z)/2 \rangle$ corresponds to an Ising order parameter. Our self-consistent mean field phase diagram is presented in Fig.~\ref{fig:2}(a). For small $J_\perp$ and $J_\parallel$, the order parameter vanishes and the ground state is a gapless $\mathbb{Z}_2$ spin liquid given by the exact solution of $H_K$. For larger $J$, we find a first-order phase transition, where the order parameter $m^z$ jumps to a finite value. Consequently, a gap ($\Delta$) emerges in the spectrum (Fig.~\ref{fig:2}(a) inset), signifying the stabilization of a $\mathbb{Z}_2$ topological order. 

This magnetic state is a ${\bf q}=0$ order, as shown in Fig.~\ref{fig:1}(b). It breaks diagonal mirror symmetry ($\mathcal{M}_{xy}$), which bisects the two red and the two green bonds of the squares (Fig.\ref{fig:1}(a)) and time reversal symmetry ($\Theta$) while it preserves their product, $\mathcal{M}_{xy} \otimes \Theta$. Since it is not possible to recover the same pattern through a combination of time-reversal symmetry and lattice translations, this state is altermagnetic \cite{Bose_PRB2024}.  To differentiate this phase from a conventional AM and the other phases discussed below, we refer to it as a topological altermagnet (tAM). Unlike standard AMs, which exhibit momentum-dependent spin-split electronic excitations, this tAM features split bands of fractionalized excitations as shown in Fig.~\ref{fig:2}(b). In the limit of $K_\perp \rightarrow 0$, these bands are marked by the layer index $A/B$. Therefore, the spin degree of freedom in a standard altermagnet is replaced by the layer index in a tAM. Since the spectrum is gapped, we present the constant-energy surface at $E=-3.2$ in Fig.~\ref{fig:2}(b). The altermagnetic splitting exhibits a d-wave form factor ($d_{x^2-y^2}$) protected by $\mathcal{M}_{xy} \otimes \Theta$ symmetry and thus the excitation spectrum remains unsplit along $\rm \Gamma - M$. 
\begin{figure}[t]
    \centering
    \includegraphics[width=\linewidth]{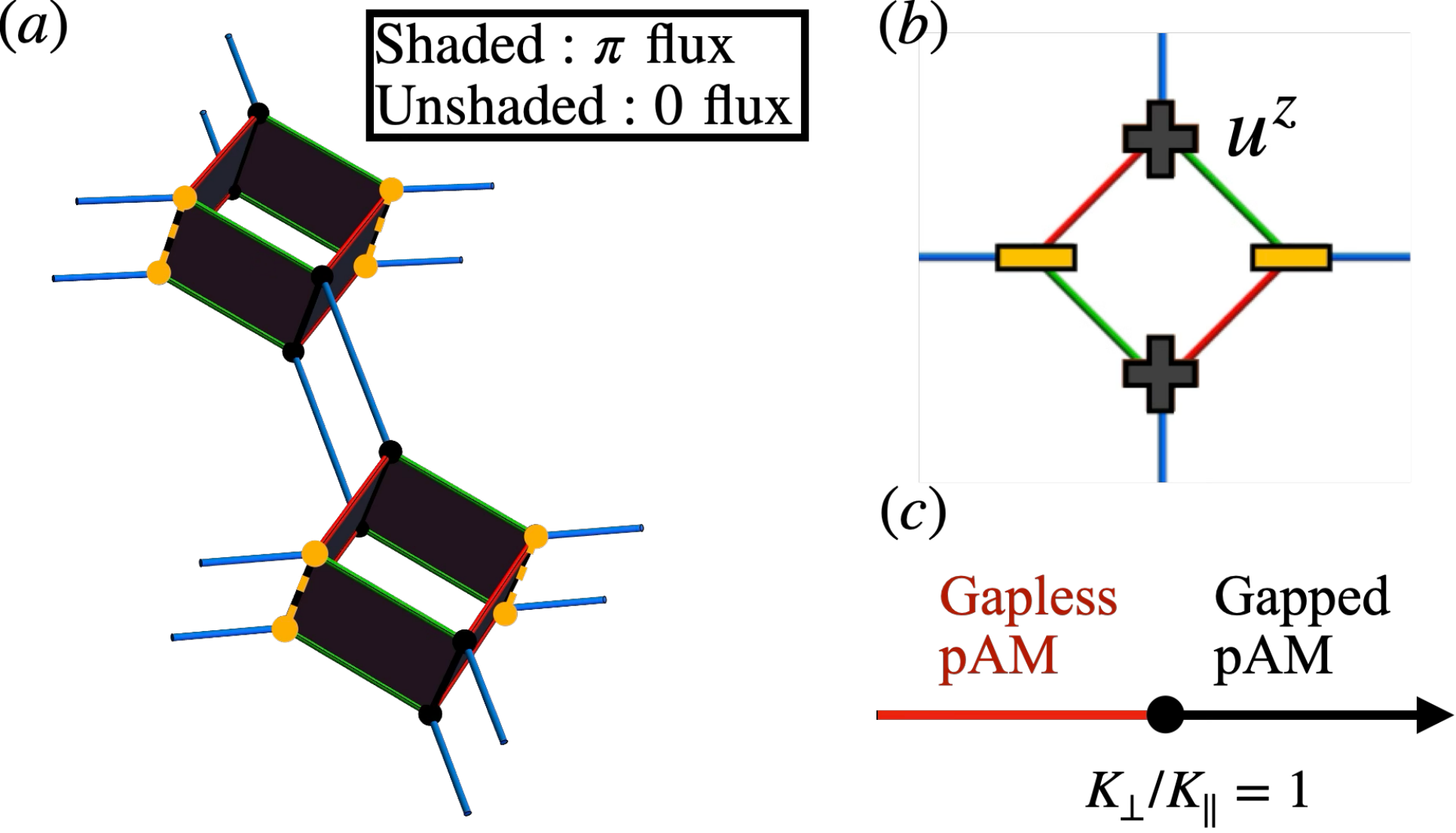}
    \caption{(a) Flux configuration choice to obtain altermagnetic splitting pAMs. (b) The gauge choice for this flux sector is achieved by choosing interlayer bonds $u^z=-1$ that are marked in yellow. (c) The phase diagram for pAMs shows a gapless to gapped transition at $K_\perp/K_\parallel=1$. }
    \label{fig:3}
\end{figure}
It should be noted that a tAM is a peculiar phase where both local and topological order coexist. Similar phases have been explored in other $\Gamma$-matrix generalizations of the Kitaev model \cite{Seifert_PRL2020, Vijayvargia_PRR2023} where the integrability is broken by the the inclusion of additional interactions while preserving the gauge structure of the spin liquid. The coexistence of a local Ginzburg-Landau type order and a nonlocal topological order is commonly referred to as `magnetic fragmentation' which was theoretically predicted \cite{Brooks_PRX2014} and experimentally observed \cite{Petit_NatPhys2016} for spin ice materials. Thus, the topological altermagnetic phase represents an example of magnetic fragmentation, where the local order is altermagnetic.

\begin{table}
\centering
\renewcommand{\arraystretch}{1.1}
\setlength{\tabcolsep}{2pt}
\small
\begin{tabular}{>{\centering\arraybackslash}m{1.3cm} *{5}{>{\centering\arraybackslash}m{1.3cm}}}
\toprule
& $\mathcal{M}_{xy}$ & $\Theta$ & $\mathcal{X}$ & $\mathcal{M}_{xy}\!\otimes\! \Theta$ & $\mathcal{M}_{xy}\!\otimes\!\mathcal{X}$ \\
\midrule
AM & \xmark & \xmark & n/a & \cmark & n/a \\
tAM & \xmark & \xmark & \xmark & \cmark & \cmark \\
pAM & \cmark & \cmark & \cmark & \cmark & \cmark \\
hAM & \xmark & \xmark & n/a & \cmark & n/a \\
\bottomrule
\end{tabular}
\caption{Summary of broken symmetries in comparison to a standard altermagnet (AM) on a square-octagon lattice. The topological altermagnet (tAM) and half-altermagnet (hAM) both break diagonal mirror symmetry $M_{xy}$ and the time reversal symmetry ($\Theta)$, while the pseudo-altermagnet (pAM) preserves all symmetries. Additionally, layer exchange symmetry ($\mathcal{X}$) is broken in the tAM phases.}
\label{tab:am_comparison}
\end{table}

{\it Pseudo-altermagnet in the absence of local order.}
The altermagnetic splitting in a tAM arises due to an effective field $m^z$ among the $f_{A/B}$ quasiparticles, $f^\dagger_\nu \sigma^z_{\nu \nu^\prime} f_{\nu^\prime}$ where $\sigma^z$ is a Pauli matrix acting on the $A/B$ subspace. Alternatively, one can explore effective field configurations aligned along different $m$ components. In particular, an effective field in the $xy$-plane couples the fermions on each layer. For instance, consider an effective field along the $x$-direction, $m^x_i = f^\dagger_{\nu i} \sigma^x_{\nu \nu^\prime} f_{\nu^\prime i} = (f_{Ai}^\dagger f_{Bi}+ f_{Bi}^\dagger f_{Ai})$ which can also act as an altermagnetic field with momentum-split bands that are eigenstates of $\sigma^x$ (in contrast to $\sigma^z$ for tAM). However, $m^{x(y)}$ changes sign under local $\mathbb{Z}_2$ gauge transformations, $m_i^{x(y)} \rightarrow - m_i^{x(y)}$, and therefore its expectation value vanishes for the physical wave function. Nevertheless, a non-local gauge invariant correlator can be defined as follows \cite{Nica_npjQM2023, Vijayvargia_PRR2023},
\begin{eqnarray}
\mathcal{C}(i,j) = \langle m_i^x B(i,j) m_j^x\rangle 
\end{eqnarray}
where the gauge string $B(i,j) = \prod_{kl \in (i, j)} u_{Akl}u_{Bkl}$ connects the operators at the end sites $(i, j)$. $\mathcal{C}(ij)$ takes the same value for all gauges and therefore also for the physical wave function. Although these arguments indicate that a local order parameter in the $xy$-plane is not possible, the presence of nonlocal correlations signifies the long-range entanglement essential for topological order.

Such an effective field naturally arises in our model through the $K_\perp$ term, which can be 
rewritten as $H_{K^\perp}= 2K^\perp\sum_{i}u^z_if_{\alpha i}^{\dag} \sigma^x_{\alpha \beta} f_{\beta i}$. Fixing the gauge to all $u^z_i=1$ in order to enforce the zero-flux sector for interlayer plaquettes results in an effective uniform $m^x$ field for all sites, like a ferromagnet.

Instead of a uniform $u^z_i=1$ bond configuration, if the $u^z_i$ bonds are arranged in an alternating $\pm1$ pattern within a unit cell, the effective field behaves as an altermagnetic field, similar to the case of a tAM. Flipping two out of the four interlayer bonds switches the interlayer plaquettes within the unit cell from 0 to $\pi$ flux, as illustrated in Fig.~\ref{fig:3}(a) and (b). This can be achieved by modifying the flux sector enforcing term to $-\lambda(\sum_p(W_{p,S}+W_{p,O})+\sum_i I_i^3)$, where the sum is only over the $\alpha=3$ blue bonds. This chemical potential term for the fluxes enforces the intralayer fluxes and interlayer fluxes on the blue bonds to be in the zero-flux state while the other two interlayer fluxes remain in the $\pi$-flux state. 

A momentum-dependent splitting can be observed in the excitation spectrum of the fractionalized quasiparticles, \textit{without}  an altermagnetic local order parameter (Fig.~\ref{fig:1}(d)). $m^x$ can be continuously tuned by $K_\perp$, resulting in a gapless phase for $|K_\perp/K_\parallel|<1$ and a gapped phase otherwise (Fig.~\ref{fig:3}(c)). The spectrum remains invariant under all gauge choices. For instance, consider a gauge transformation where $u_{B ij}=-1$ for the green and red bonds (bonds that are within the bottom layer squares), while all the other bond variables remain $u_{A ij}=1$ and all $u^z_i=1$.  This transformation preserves the flux sector; however, the Hamiltonian can no longer be expressed as an altermagnetic local field, even though the spectrum continues to exhibit momentum-dependent splitting.

We refer to this phase as a pseudo-altermagnet (pAM) since, despite its resemblance to a tAM in terms of excitation spectrum, the absence of a local AM order prevents the key physical properties of AMs from being observed. Specifically, while there appears to be a breaking of $\mathcal{M}_{xy}$ in the initial gauge depicted in Fig. \ref{fig:3}(b), the aforementioned gauge transformation restores $\mathcal{M}_{xy}$. Since the physical wave function is gauge-symmetrized, it does not break any spatial or time-reversal symmetry, confirming the absence of a local order parameter.

\begin{figure}
    \centering
    \includegraphics[width=0.9\linewidth]{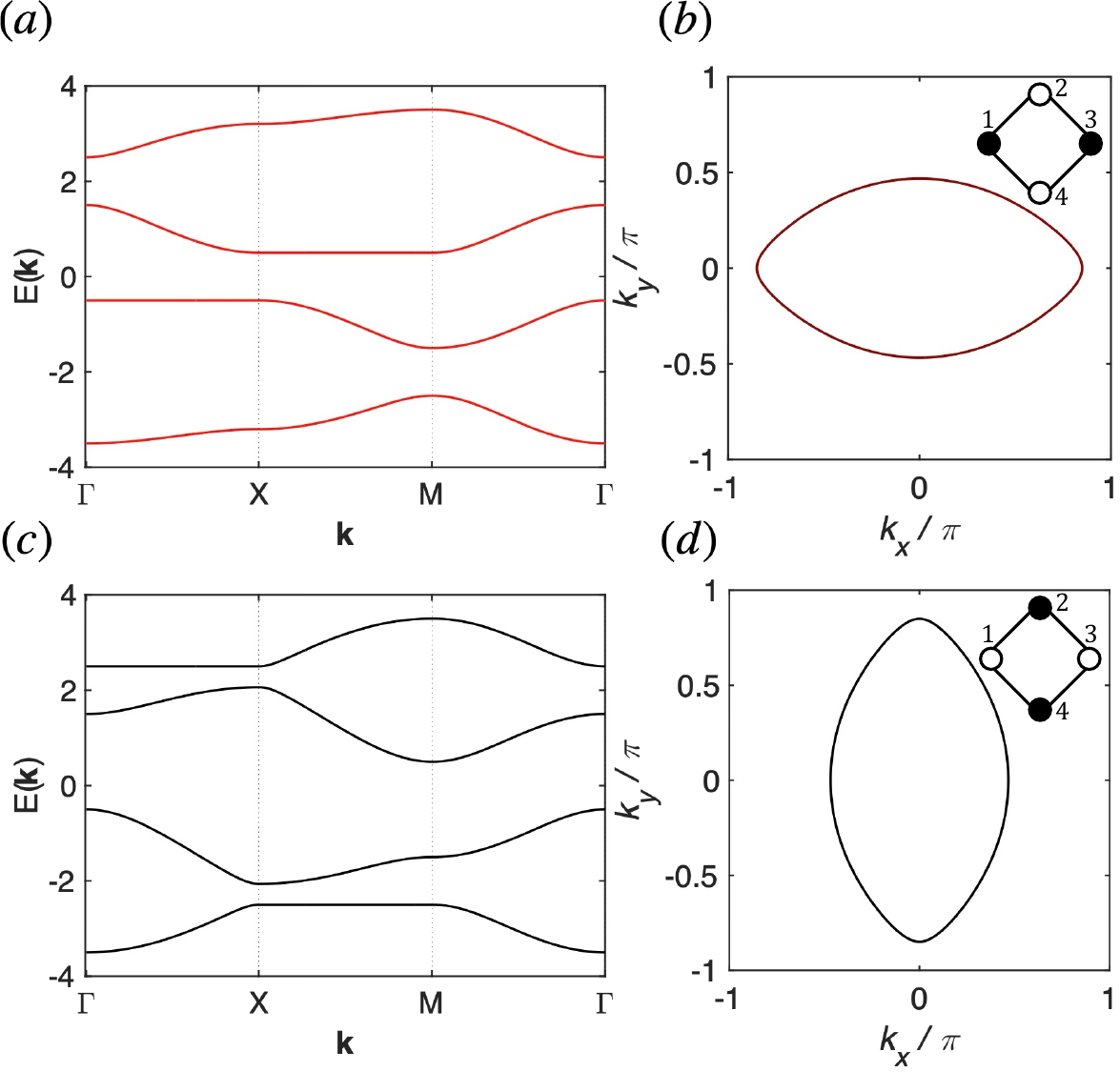}
    \caption{Monolayer with half-altermagnetic order. (a), (c) The excitation spectrum and (b), (d) the constant energy surface at $E=-3.2$ for two types of AM spin configuration given as shown in the inset.
}
    \label{fig:4}
\end{figure}

{\it Half-altermagnets in monolayers.}
Next, we examine the monolayer limit of tAMs (see SM for details). Restricting our analysis to the zero-flux sector, our mean-field solutions correspond to the $y$-axis of Fig. \ref{fig:2}(a). In the magnetically ordered phase, a gap opens, and the order parameter spontaneously breaks the effective $C_4$ symmetry of the complex fermion Hamiltonian down to $C_2$, as well as breaking diagonal mirror symmetry ($\mathcal{M}_{xy}$) symmetry. Within a unit cell, two possible magnetic configurations emerge, as illustrated in Fig. \ref{fig:4}. These configurations are related by either time-reversal symmetry or $\mathcal{M}_{xy}$. Due to spontaneous symmetry breaking, the ground state selects one of these configurations, resulting in an excitation spectrum with only a single band, depending on the chosen configuration. Since this phase possesses only half the DOF of a tAM, we refer to it as a half-altermagnet (hAM), drawing an analogy to half-metals.

{\it Experimental signatures.} There are several experimental signatures of standard altermagnets including anomalous Hall effect, spin current, and Kerr effect. Although these responses are commonly attributed to materials with net magnetization, AMs show similar behavior due to spin-split bands\cite{PhysRevLett.128.197202,PhysRevLett.130.216702,PhysRevLett.132.056701,liao2024separation,feng2022anomalous,cui2023efficient,sun2023spin,bai2023efficient,gray2024time}. 
In Table~\ref{tab:am_comparison}, we present the symmetries of our three phases compared to a conventional AM. It is clear that the topological altermagnetic phase has a local order that breaks the same symmetries as an ordinary altermagnet, which would lead to the same physical observables. However, due to the nature of the $\mathbb{Z}_2$ spin liquid in this system, transport quantities come from the fractionalized excitations $f_{A/B}$ rather than electronic quasiparticles. These excitations do not carry charge, so Hall effect measurements will not be possible. However, they carry entropy, layer index, and $S^z$ polarization, as their density is gauge invariant and proportional to $\sigma^z_{A/B} = 2n_{A/B}-1$. So, the $f$ current can carry both spin \cite{Keskiner_arXiv2025} and heat. 
As a result, layer-dependent thermal and $S^z$ transport are key signatures for detecting tAMs. Similarly, hAMs can exhibit anisotropic thermal and spin transport as they have only a single component of $f$ quasiparticles. However, pAMs do not exhibit these properties since they have no local order, even though their dynamical spin structure will be similar to tAMs when the layer DOF are summed over.

We developed a quantum spin model that incorporates the essential aspects of altermagnetism and Kitaev spin liquids. We showed that the altermagnetic order can naturally arise when there is sufficient DOF among itinerant quasiparticles. The excitations of tAMs involve fractionalized quasiparticles that are split in momentum space. Additionally, we propose experimental signatures for detecting these phases. Interesting future directions include generalizing our model to U(1) spin liquids and exploring moir\'e superlattices of tAMs.

{\it Note added}: During the completion of this work, we learned of a forthcoming work \cite{Urban_unpub} on a related problem.

We thank Nandini Trivedi and Turan Birol for fruitful discussions. AV and OE acknowledge support from NSF Award No. DMR-2234352. EDR and ABS acknowledge support from NSF Award No. DMR-2206987. We thank the ASU Research Computing Center for high-performance computing resources.

\bibliography{references.bib}
\end{document}

\title{Supplementary material for ``Altermagnets with topological order in Kitaev bilayers"}
\author{Aayush Vijayvargia, Ezra Day-Roberts, Antia S. Botana, Onur Erten}
\affiliation{Department of Physics, Arizona State University, Tempe, AZ 85287, USA}

\maketitle

\section{Gauge choice}
\begin{figure}[!ht]
    \centering
    \includegraphics[width=0.25\linewidth]{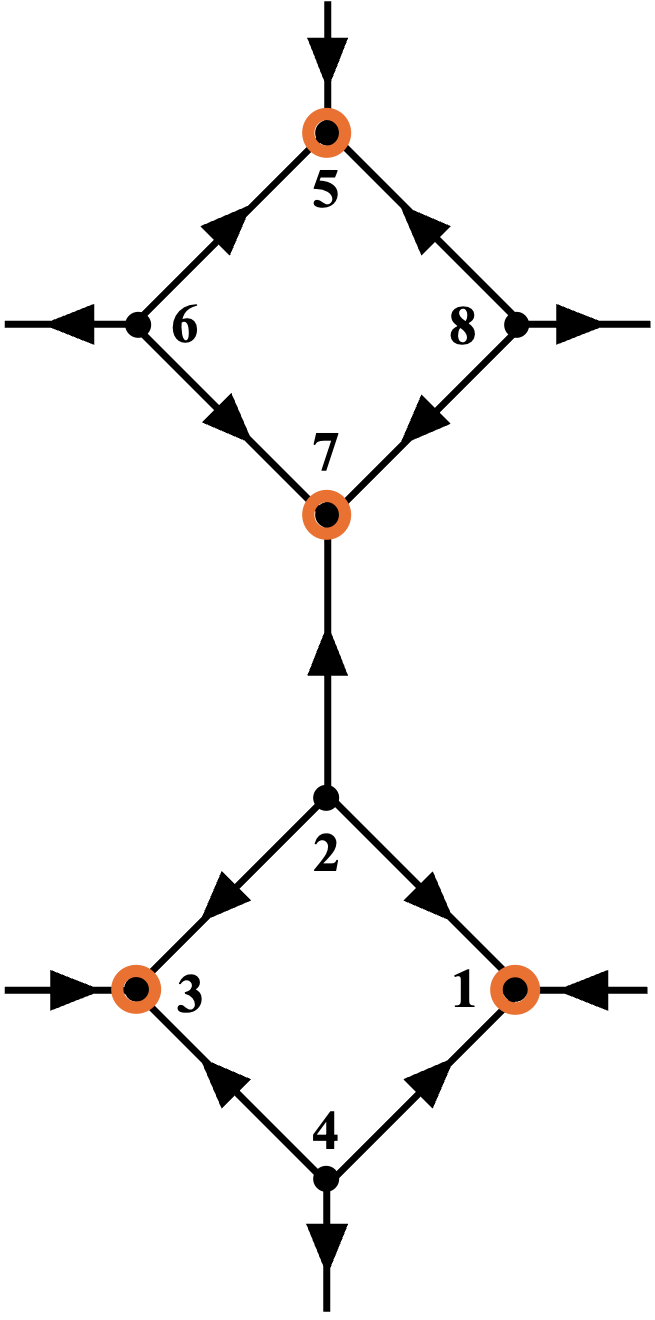}
    \caption{Choice of gauge for the fermion bands. Sites labeled by an orange ring are the site of gauge transformations. Bond variables $u_{\nu ij}$ are directed quantites, with the direction labeled for each bond.}
    \label{fig:5}
\end{figure}

\noindent The non-interacting part of the Hamiltonian in the complex fermion representation is given by
\begin{align}
     H=& 2K\sum_{\langle ij \rangle,\nu}u_{\nu,ij}(\text{i}f_{\sigma i}^{\dag}f_{\nu j}+ {\rm H.c.})+2J_\perp\sum_{i}u^z_i(\text{i}f_{A i}^{\dag}f_{B i}+ {\rm H.c.})\nonumber
\end{align}
We extend our model to have a larger unit cell as shown in Fig. \ref{fig:5}. Next, we perform a staggered gauge transformation on alternating sites, marked by a ring around the site: $f_{\nu i}\rightarrow\text{i}f$. This resultingly makes the hopping coefficients real, as depicted in Eq. 5.
Furthermore, the intralayer bond variables $u_{\nu,ij}=-u_{\nu,ji}$. Therefore, we follow a definite ordering of the bond variable indices as depicted by the arrows in Fig. \ref{fig:5}. In the zero-flux sector, $u_{\nu,ij}=1$ along a bond  pointing from $i\rightarrow j$.
\section{Dangling Majorana fermions in hAMs}
In the bilayer model, there are four bonds emanating from a site and consequently four bond variables. However, in the monolayer limit, this treatment leaves one Majorana fermion $c^z_i$ per site to be dangling. These Majorana fermions can be combined along all the $\alpha=3$ blue bonds, and a chemical potential term can be added to the Hamiltonian to move the band away from the Fermi energy of the system. 
We define $f^z_{i}=c^z_i+ic^z_{i+\delta_z}$. Further we add $\mu_z\sum_if^{z \dag }_if^z_i$, with $u_z>0$, to ensure that this Majorana fermion band is gapped.